\documentclass[iop,apj]{emulateapj}
\usepackage{amsmath}

\usepackage{graphicx,natbib,amsmath,amssymb}
\bibpunct[, ]{(}{)}{;}{a}{}{,}
\newcommand{\simgt}{\hbox{\rlap{\raise 0.425ex\hbox{$>$}}\lower 0.65ex\hbox{$\sim$}}}
\newcommand{\simlt}{\hbox{\rlap{\raise 0.425ex\hbox{$<$}}\lower 0.65ex\hbox{$\sim$}}}

\defcitealias{2010ApJ...722..851H}{Paper I}
\defcitealias{2010ApJ...722..856W}{Paper II}
\defcitealias{2010ApJ...725..282W}{Paper III}

\journalinfo{The Astrophysical Journal, accepted (2014)}
\submitted{Received 2013 August 12; accepted 2014 January 8}

\shorttitle{DARKexp: Distribution of angular momentum}
\shortauthors{Williams et al.}

\begin{document}

\title{
Statistical mechanics of collisionless orbits. 
IV. Distribution of angular momentum}


\author{ 
Liliya L. R. Williams\altaffilmark{1},
Jens Hjorth\altaffilmark{2}, and
Rados{\l}aw Wojtak\altaffilmark{2}
}

\altaffiltext{1}{School of Physics and Astronomy, University of Minnesota, 
116 Church Street SE, Minneapolis, MN 55455, USA; llrw@astro.umn.edu}
\altaffiltext{2}{Dark Cosmology Centre, Niels Bohr Institute, University of 
Copenhagen, Juliane Maries Vej 30, DK--2100 Copenhagen \O, Denmark;
jens@dark-cosmology.dk,wojtak@dark-cosmology.dk}


\begin{abstract}
It has been shown in previous work that DARKexp, which is a theoretically 
derived, maximum entropy, one shape parameter model for isotropic collisionless 
systems, provides very good fits to simulated and observed dark-matter halos. 
Specifically, it fits the energy distribution, $N(E)$, and the density 
profiles, including the central cusp. Here, we extend DARKexp $N(E)$ to 
include the distribution in angular momentum, $L^2$, for spherically symmetric 
systems. First, we argue, based on theoretical, semi-analytical, and 
simulation results, that while dark-matter halos are relaxed in energy, 
they are not nearly as relaxed in angular momentum, which precludes using 
maximum entropy to uniquely derive $N(E,L^2)$. 
Instead, we require that when integrating $N(E,L^2)$ over squared angular momenta
one retrieves the DARKexp $N(E)$. Starting with a general expression for 
$N(E,L^2)$ we show how the distribution of particles in $L^2$ is related to 
the shape of the velocity distribution function, VDF, and velocity anisotropy profile, 
$\beta(r)$. We then demonstrate that astrophysically realistic halos, as 
judged by the VDF shape and $\beta(r)$, must have linear or convex 
distributions in $L^2$, for each separate energy bin. The distribution in energy of 
the most bound particles must be nearly flat, and become more tilted in favor of 
radial orbits for less bound particles. These results are consistent with
numerical simulations and represent an important step towards deriving
the full distribution function for spherically symmetric dark-matter halos.
\end{abstract}





\section{Introduction}

The full dynamical description of relaxed dark-matter halos is of fundamental 
importance for our basic understanding as well as for more practical 
applications in galaxy formation and evolution, and cosmology. N-body 
simulations have converged on the properties of dark-matter halos 
\citep{2004MNRAS.349.1039N,2009MNRAS.398L..21S,2010MNRAS.402...21N}, though 
some uncertainty may still remain arising from finite resolution effects.

A lot of work has been devoted to attempts to explain the density and velocity 
structure of relaxed halos by examining the dynamical processes at work, like 
mass accretion rate, conservation of radial action, radial orbit instability, 
etc.\ 
\citep[e.g.,][]{2003A&A...408...27L,2004ApJ...604...18W,2006MNRAS.368.1931L,2007ApJ...666..181S,2007MNRAS.376..393A,2010arXiv1010.2539D}. 
Though such phenomenological arguments are valuable and bring insight to the 
problem, they do not arise directly from fundamental physics. Because the 
properties of virialized dark matter halos appear universal, and are only 
weakly dependent on initial conditions, like cosmological model, local 
density, etc., it is reasonable to assume that the structure of halos is 
governed by physics more fundamental than that described by phenomenology.

Motivated by the possibility of a first principles solution, several groups 
have attempted a statistical mechanics approach. To our knowledge, the first 
attempts were made by \citet{1957SvA.....1..748O} and \citet{1967MNRAS.136..101L}, 
with somewhat limited success 
\citep[see][hereafter \citetalias{2010ApJ...722..851H}, for a discussion]{2010ApJ...722..851H}. 
Several other works have appeared since 
\citep{1987MNRAS.229...61S,1991MNRAS.253..703H,1992ApJ...397L..75S,1998MNRAS.300..981C,2008PhRvE..78b1130L,2009ApJ...692..174L,2013MNRAS.430..121P}.
Some of these rely on arbitrary assumptions, while others require many 
parameters to fit halos adequately.

Our statistical mechanics approach (\citetalias{2010ApJ...722..851H}) differs 
from the previous ones in that we work in a different state space, which we 
argue is more appropriate for collisionless systems, and uses an accurate 
description of low occupation numbers, which appears to be important for 
self-gravitating systems \citep[see also][]{1996MNRAS.280.1089M}. In 
\citetalias{2010ApJ...722..851H} we derived the differential mass, or energy 
distribution, $dM/dE=N(E)$ for collisionless material under the 
assumption that the final steady-state configuration represents the most likely
state, and therefore can be obtained as a maximum entropy state. Our model is
called DARKexp, and its energy distribution is $N(\epsilon)=\exp(\phi_0-\epsilon)-1$,
where $\epsilon=\tilde\beta E$ is the dimensionless energy, $\tilde\beta$ is the inverse 
thermodynamic temperature ($\tilde\beta<0$), and $\phi_0=\tilde\beta \Phi_0$ is the 
dimensionless central potential. 

Unlike other theoretically motivated density profiles 
\citep[][]{1966AJ.....71...64K,1967MNRAS.136..101L,1996MNRAS.280.1089M}, 
DARKexp predicts central density cusps. The asymptotic, small $r$ density 
slope for all central potentials, $\phi_0$, is $d\ln\rho/d\ln r=-1$, but for 
radii accessible to N-body simulations, the central slope varies depending on 
$\phi_0$: $\phi_0\,\simlt\, 4$ systems have inner slopes shallower than $-1$, while 
$\phi_0\,\simgt\, 5$ have inner slopes between $-1$ and $-2$.

DARKexp appears to be a very good descriptor of the energy distribution and 
density profile of dynamical systems: galaxy and galaxy cluster size 
dark-matter halos in simulations
\citep[][hereafter \citetalias{2010ApJ...725..282W}]{2010ApJ...725..282W},
observed galaxy clusters \citep{2013arXiv1301.1684S}, and even many globular 
clusters \citep{2012MNRAS.423.3589W}. \citet{2013arXiv1301.1684S} compared 
a range of theoretical and phenomenological models to relaxed galaxy clusters
whose profiles were estimated using strong and weak lensing. DARKexp did 
better than other theoretical models, and performed as well as the best 
empirical fitting functions.

The DARKexp model derived in \citetalias{2010ApJ...722..851H} has one 
limitation: it describes the distribution of particles in energy only, i.e.,
$N(E)$, and implicitly assumes that the distribution in angular momentum 
corresponds to that of a system with an isotropic velocity ellipsoid. While 
velocity anisotropy (hereafter, anisotropy) affects the shape of $N(E)$ only 
weakly, one still would like to know the full dynamical description of a system, 
including the distribution of particles in angular momentum, $L$. Systems 
considered in this work have no net rotation, so the angular momentum
vector is reduced to its modulus. Dark-matter halos in simulations, and 
probably in the Universe are not isotropic. Therefore one needs to extend 
DARKexp $N(E)$ to include $L^2$.  

This paper is devoted to estimating $N(E,L^2)$ from basic arguments and 
simple models. In Section~\ref{genL2} we discuss the degree of mixing
in energy and angular momentum and argue that a maximum entropy approach
may not be applicable to the problem in hand. Instead we propose an
integral constraint on $N(E,L^2)$. In Section~\ref{secIII} we generate
and characterize halos obeying this constraint and compare to simulated
halos. Section~\ref{secIV} provides a summary and an outlook.

\section{General considerations for $N(E,L^2)$}\label{genL2}

\subsection{The lack of mixing of angular momenta}

Relaxation into equilibrium, or at least into a long-lived steady state can 
be driven by any dynamical process that mixes, or redistributes, particle 
energies and angular momenta, thereby causing them to be uncorrelated with the 
corresponding initial values. Dark-matter halos in simulations are nearly 
relaxed systems, though the degree of relaxation is still not clear. In a 
collisionless system, mixing in energy is achieved because the particles exchange 
energy with the global time-varying potential \citep{1967MNRAS.136..101L}, while 
mixing in angular momentum is accomplished through torques which are present 
in any system that deviates from spherical symmetry.

We now argue that dark-matter halos are not as well relaxed in angular momentum 
as they are in energy. This difference in the degree of relaxation in the two 
parameters, $E$ and $L^2$, is possible because relaxation in energy and in 
angular momentum can proceed relatively independently of each other, at least
in some systems. One example is the Extended Secondary Infall Model (ESIM), 
described in 
\citet[][hereafter \citetalias{2010ApJ...722..856W}]{2010ApJ...722..856W}
and some earlier works \citep{1987ApJ...318...15R,2004ApJ...604...18W}. 
ESIM systems relax through spherically 
symmetric collapse in which particles/orbits keep their initial $L^2$ 
throughout the evolution. Even though $L^2$ of individual particles do not 
change at all, the corresponding $E$ does change significantly, and the final 
virialized halos are well fit with DARKexp. Therefore mixing and relaxation 
in $E$ can be achieved, without having any mixing in $L^2$. 

\begin{figure}
\includegraphics[width=18cm,height=18cm,keepaspectratio]{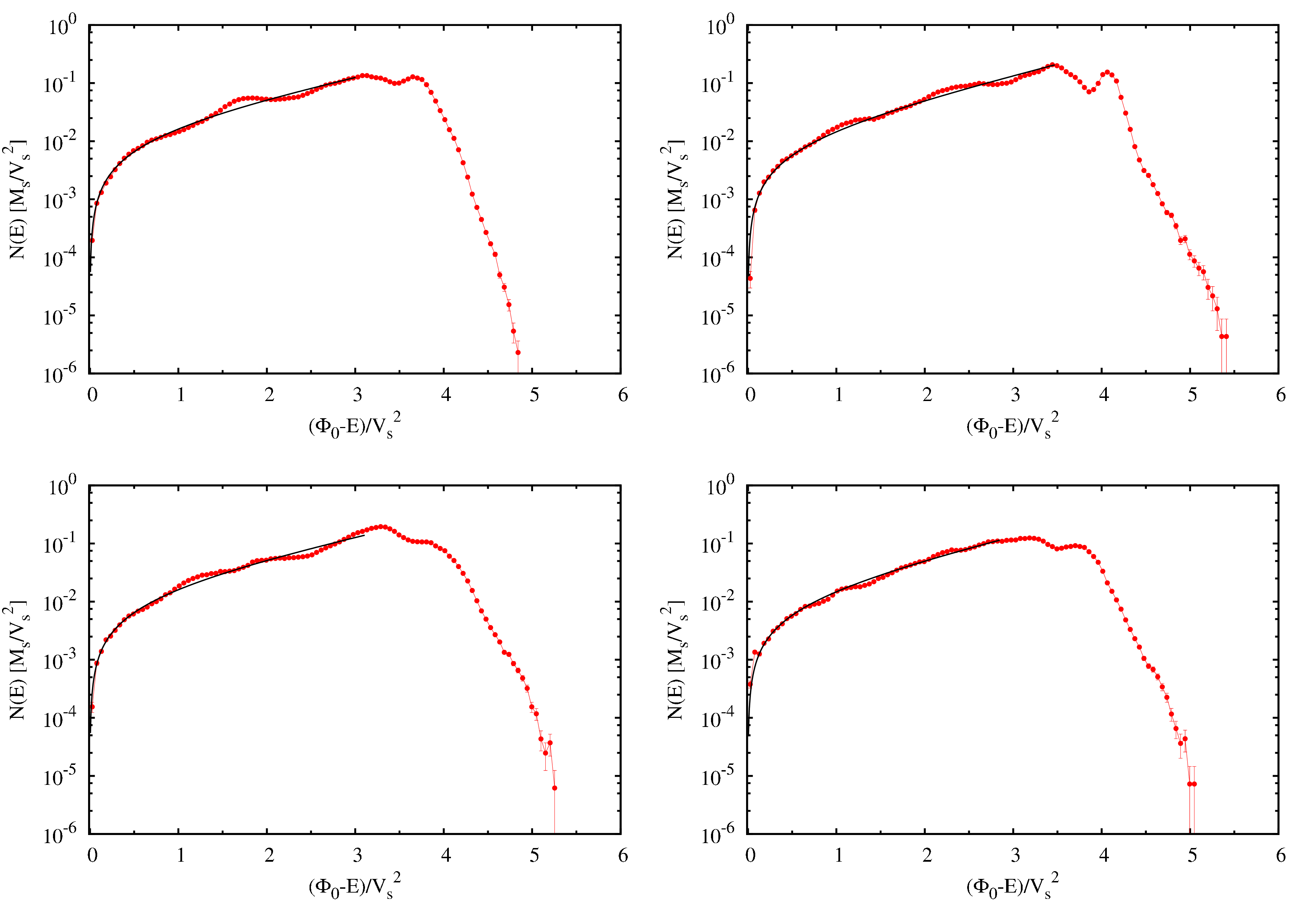}
\caption{Comparison between $N(E)$ of four randomly chosen halos from a cosmological $N$-body simulation 
(red curves with points and Poisson errorbars) and the DARKexp fits (black curves). The scalings on the 
axes are in terms of parameters ($M$ is the enclosed mass, $V$ is the circular velocity) measured at 
$r_{-2}$, which is the radius where the logarithmic density slope, $d\ln \rho(r)/d\ln r$, is equal to $-2$; 
see Section 3.1 of \citetalias{2010ApJ...725..282W} for details. 
}
\label{DEind}
\end{figure}

Numerically simulated dark-matter halos in equilibrium are relaxed in energy 
as evidenced by their being well fit with DARKexp 
(\citetalias{2010ApJ...725..282W}). Figure~\ref{DEind} contains four individual
halos and their DARKexp fits that make up the average shown in that paper. The fit
at very large negative energies (i.e. for very bound particles) and intermediate energies 
is very good; at small negative energies (right side of each panel) most of the 
particles are quite far from the halo center, and so may not be in equilibrium. 

Are the halos equally well relaxed in $L^2$? 
Apparently not. \citet{2013MNRAS.434.1576W} show that the principal axes of 
cosmologically simulated halos are aligned with the {\em local} velocity 
ellipsoids, and the alignment is strongest in the innermost shells. Moreover,
the principal axes are aligned with the large scale structure filaments 
\citep{2013MNRAS.428.2489L} which
implies that the angular distribution of angular momentum, $\boldsymbol L$, retains 
the memory of the formation process, down to the inner most regions. If mixing 
in angle were complete, as required by full relaxation, these alignments 
should have been erased. 

Why are halos not well mixed in $L^2$? We speculate that this is because in 
a collapsing system, radial forces, and hence changes in radial forces, tend 
to be larger and longer lasting than tangential ones. The former are largely 
responsible for mixing in $E$, while the later are exclusively responsible for
mixing in $L^2$. In other words, radial fluctuations in the gravitational 
potential are more dominant than tangential ones, where the latter are brought 
about by ellipticity, substructure and merging. 

\subsection{Failure of maximum entropy arguments for the angular momenta}

As a consequence, this observation implies that theoretical maximum entropy 
approaches that assume full mixing in angle and require $L^2$ to be part of 
the entropy \citep[e.g.,][]{2013MNRAS.430..121P}, likely cannot capture the 
properties of simulated dark-matter halos.
To check that free redistribution or mixing of $L^2$ is not taking place in 
collapsing systems, we considered the final state of a system assuming that 
it does. In other words, we apply a maximum entropy argument to $L^2$ as well
as to $E$. To do this we extend our derivation presented in 
\citetalias{2010ApJ...722..851H} to include angular momentum. As in that paper, 
a maximum entropy procedure is applied, where in addition to the total energy 
and total mass, a quantity relating to the total angular momentum is also held 
fixed. Apart from the fact that there are now three, rather than two, Lagrange 
multipliers, the maximization of entropy procedure is the same as the one used 
to derive the isotropic DARKexp. The final derived distribution has the form
\begin{equation}
N(E,L^2)\propto \exp(\tilde\beta\Phi_0-\tilde\beta\,E-\gamma\,L^\xi)-1, \label{NI}
\end{equation}
which is analogous to Equation~(30) of \citet{1994ApJ...424..106H}.
(Note that $\tilde\beta$ in the above is not to be confused with the velocity anisotropy.
We denote the velocity anisotropy by $\beta$ or $\beta(r)$.)

In general, there is no analytical way to derive the density profiles from Equation~(\ref{NI}). 
We therefore used an iterative procedure, similar to the one 
used in \citetalias{2010ApJ...722..856W}, to obtain the corresponding density 
profiles. We experimented with several combinations of parameters 
$\gamma$ and $\xi$ (see Appendix~\ref{appA} for details). Some parameter 
combinations did not produce viable halos, i.e., did not converge. Those 
that did converge had velocity anisotropy profiles that were isotropic at small radii and 
became tangentially anisotropic at large radii. Because no dynamically 
produced collisionless systems is known to have such anisotropy, and because it 
seems unlikely that any dynamical process would lead to such a system, we 
conclude that systems described by Equation~(\ref{NI}), with $\gamma\ne 0$, 
do not exist in simulations, or the real Universe. 

One could argue that the choice of a power law form for $L$ in 
Equation~(\ref{NI}) is limiting, and there could be other functional forms 
that would produce realistic systems. While possible, we speculate that the
general behaviour of solutions will be similar to that displayed by 
Equation~(\ref{NI}), regardless of the what function of $L$ is used. A more 
thorough investigation is needed to explore this question.

Though Equation~(\ref{NI}) cannot be solved analytically to produce $\rho(r)$
in general, in Appendix~\ref{appB} we consider a special case that can be 
solved analytically. In this case the density profile is
\begin{equation}\label{rhoME}
\rho(r)=\frac{1}{4\pi}\Bigl(-\frac{\tilde\beta}{\gamma\xi}\Bigr)^{2/(\xi-2)}\Bigl(\frac{2+\xi}{2-\xi}\Bigr)\,r^{-4(\xi-1)/(\xi-2)},
\end{equation}
where $\tilde\beta<0$, $\gamma<0$, and $-2<\xi<0$. Equation~(\ref{rhoME}) is a power 
law, ranging in slope between $\rho\propto r^{-2}$ and $\rho\propto r^{-3}$. 
These are grossly inconsistent with the profiles obtained in simulations,
whose slopes are not constant, and steepen with radius from around $-1$ to $-3$ well within the virial 
radius \citep[e.g.,][]{1997ApJ...490..493N}.

Because Equation~(\ref{NI}) does not seem to produce systems with realistic 
density profiles and constant or increasing velocity anisotropy profiles, we 
conclude that the maximum entropy argument cannot be used to derive the energy 
and angular momentum distribution of dynamically evolved collisionless systems.
The combined evidence of the arguments presented above leads us to conclude 
that dynamically evolved collisionless systems are not as well relaxed in 
angular momentum as they are in energy; in other words, the particle angular 
momenta $\boldsymbol {L}$, and hence their moduli $L$ are not as freely redistributed 
during evolution as their energies.


\subsection{Proposed integral form for $N(E,L^2)$}

Our hypothesis that mixing in $E$ is achieved relatively quickly and 
efficiently, while mixing in $L^2$ is not, suggests that the general form 
for $N(E,L^2)$ should be
\begin{equation}
N_{\rm{DARKexp}}(E)=\int_0^{L_{\rm{max}}^2(E)}\,N(E,L^2)\,dL^2.\label{NII}
\end{equation}
Here $N(E,L^2)$ is non-separable, and upon integration over all angular 
momenta gives DARKexp $N(E)$. This means that systems that have DARKexp 
$N(E)$ can have a range of $L^2$ distributions. The different types of $L^2$ 
distributions can, for example, come about as a consequence of different
formation scenarios, such as cosmological vs.\ isolated collapse.

Equation~(\ref{NII}) is also consistent with an unrelated property of 
differential energy distributions in general. Differential energy 
distributions, $N(E)$, depend primarily on the density profile, with little 
dependence on the velocity anisotropy. The velocity anisotropy is defined
in the usual way, $\beta(r)=1-[\sigma_\theta(r)/\sigma_r(r)]^2$, where $\sigma_\theta$ 
and $\sigma_r$ are velocity dispersions in one of the tangential directions, 
and radial direction, respectively. Anisotropy is 
determined by the distribution of particles/orbits in $L^2$. This means that 
a given $\rho(r)$,  coupled with different forms for $\beta(r)$, will result 
in nearly the same $N(E)$. The insensitivity of $N(E)$ to velocity anisotropy 
was pointed out by \citet{2008gady.book.....B} (their Section 4.4 and 
Figure 4.15b) using isotropic and fully radially anisotropic Jaffe models. 

This lack of sensitivity of $N(E)$ to the velocity anisotropy ensures that 
DARKexp density profile, derived for isotropic orbits, should also describe 
anisotropic cosmological N-body simulated halos, as was shown explicitly in 
\citetalias{2010ApJ...725..282W}. The recent finding by \cite{2013arXiv1301.1684S} 
that DARKexp density profiles provide very good fits to observed galaxy clusters, 
whose galaxies \citep{2009A&A...501..419B,2013arXiv1307.5867B} and dark matter 
\citep{2009ApJ...690..358H} appear to have radial velocity anisotropy at large 
radii, is also consistent with the energy distribution being largely independent 
of anisotropy. 

In Section~\ref{secIII} we will use Equation~(\ref{NII}) to arrive at the full distribution of particles in $E$ and 
$L^2$, or $N(E,L^2)$.


\section{Distribution of $L^2$ in halos}\label{secIII}

We generate as wide a range of $N(E,L^2)$ distributions as possible, all 
satisfying  Equation~(\ref{NII}). $N(E)$ is identical for all systems, and 
the density profiles are similar for all halos (DARKexp $\phi_0=4$). The
major difference between systems is in the velocity dependent properties, 
which are related to the $L^2$ distribution. We then ask how the distribution 
of particles in angular momentum relates to the astrophysically relevant halo 
quantities, namely the velocity distribution function, VDF, and the anisotropy 
profile, $\beta(r)$. Here we will mostly deal with the radial VDF, which is a 
histogram of radial speeds of particles (with respect to the halo center) in a 
specified radial range within the halo.
 
In order to find a relation between $N(E,L^2)$, the VDF and $\beta(r)$, we 
need to quantify these properties in a succinct way, such that, for example, 
the entire $\beta(r)$ profile is represented by a single value. We do this in 
Section~\ref{charac}.  Even though all our halos are physically possible, not 
all are astrophysically realistic. For example, an astrophysically realistic 
halo cannot have large (spherically averaged) $\beta(r)$ at small radii. We then 
isolate $N(E,L^2)$ distributions that give rise to astrophysically realistic 
VDFs and anisotropy profiles. 

\subsection{Generating the $L^2$ distributions}\label{caseIII}

The general method for generating $N(E,L^2)$ uses Equation~(\ref{NII}). We 
start with the DARKexp form for $N(E)$, and at each energy distribute 
particles in $L^2$ according to some prescription. (Same method as used in 
\citetalias{2010ApJ...725..282W}.) We work with dimensionless energy units, 
$\epsilon$, defined in \citetalias{2010ApJ...722..851H}, and 
$\ell^2=L^2/L_{\rm{circ}}^2$, where $L_{\rm{circ}}(\epsilon)$ is the angular momentum for 
a circular orbit at that energy. We use a total of eight different prescriptions.
Here is one example:
\begin{equation}\label{ex1}
\begin{split}
N(\epsilon,\ell^2)=N_{\rm{DARKexp}}(\epsilon)\, 
\Bigl[{1-\frac{a}{c+1}{\Bigl(\frac{\phi_0-\epsilon}{\phi_0}\Bigr)}^b}\Bigr]^{-1} \\ 
\times\Bigl[1-a\Bigl(\frac{\phi_0-\epsilon}{\phi_0}\Bigr)^b\Bigl(\ell^2\Bigr)^c\Bigr].\quad
\end{split}
\end{equation}
The first square bracket contains the normalization factor, which depends on energy,
while the second square bracket contains the dependence on $L$ and energy. This makes
$N(E,L^2)$ non-separable (as is true for all our prescriptions).  Another example is
\begin{equation}\label{ex2}
\begin{split}
N(\epsilon,\ell^2)=N_{\rm{DARKexp}}(\epsilon)\,\Bigl[b^{-1}\,c^{1/b}\Bigl(\frac{\phi_0-\epsilon}{\phi_0}\Bigr)^{-a/b}\\
\times \int_0^{t_{\rm{circ}}}\,e^{-t}\, t^{1/b-1}\,dt\Bigr]^{-1}\,e^{-t}, \\
t=c^{-1}\,\Bigl(\frac{\phi_0-\epsilon}{\phi_0}\Bigr)^a\,\ell^{2b},
\end{split}
\end{equation}
where the expression in the square brackets is the normalization factor, and 
is proportional to the lower incomplete gamma function, with $t_{\rm{circ}}$ being 
$t$ when $L=L_{\rm{circ}}$, or $\ell=1$. Note that constants $a$, $b$, and $c$ in 
Equations~(\ref{ex1}) and (\ref{ex2}) are not the same. Different realizations 
of Equations~(\ref{ex1}) and (\ref{ex2}) use a range of values for these 
constants. 

\begin{figure}
\includegraphics[width=16cm,height=16cm,keepaspectratio]{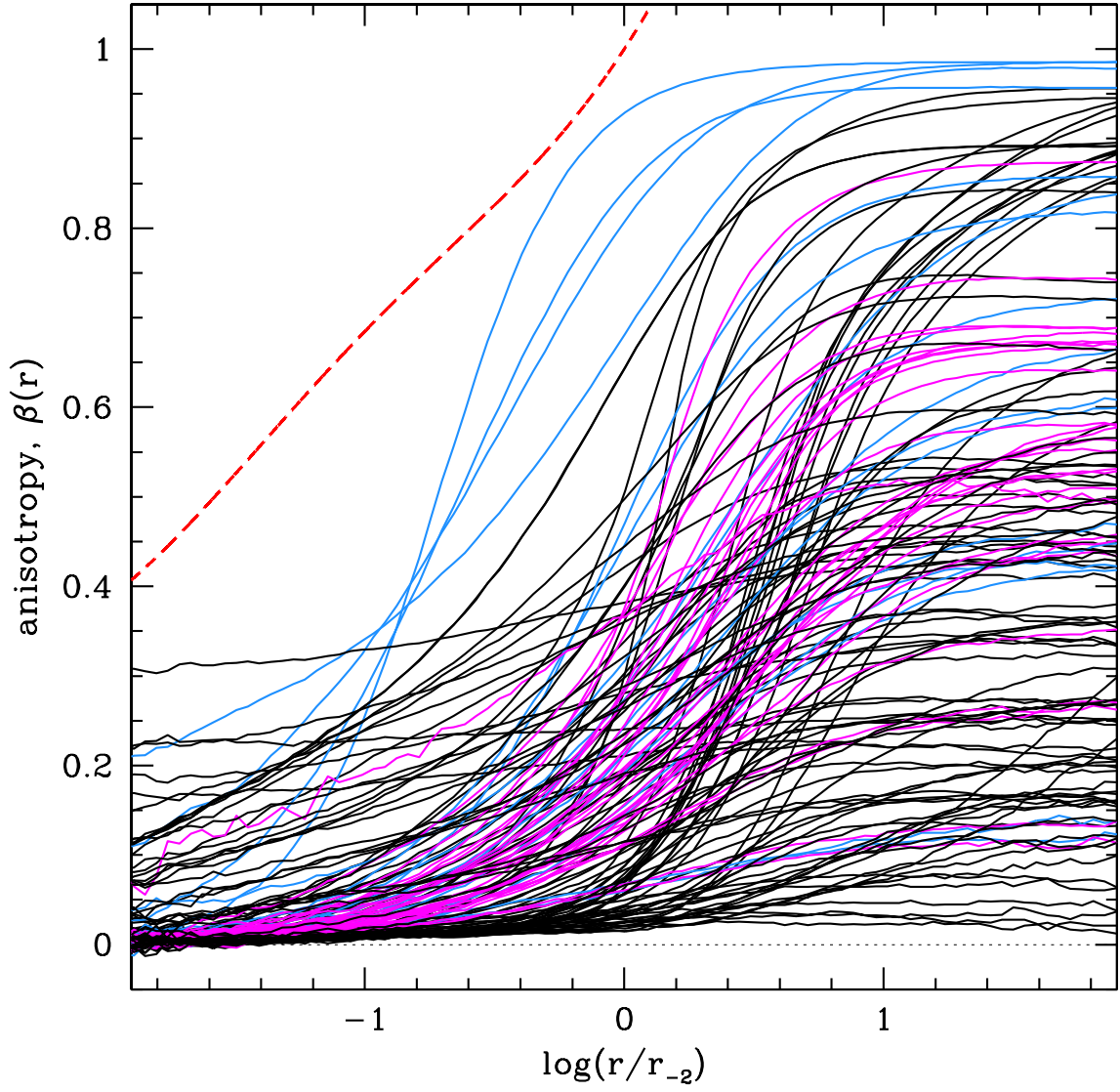}
\caption{Anisotropy profiles for all the halos used in this paper. The red dashed curve 
corresponds to $\frac{1}{2}\,[-d\ln \rho(r)/d\ln r]$ of DARKexp $\phi_0=4$, and represents 
the upper limit on the anisotropy derived by \citet{2006ApJ...642..752A,2010MNRAS.408.1070C}.
We disallowed monotonically decreasing values of $\beta(r)$. Small fluctuations 
in $\beta(r)$ are due to numerical noise. We highlight in color $\beta(r)$ profiles from
two prescriptions: Equation~\ref{ex1} and \ref{ex2} are shown as magenta and blue curves,
respectively.
}
\label{betaprof}
\end{figure}

\begin{figure}
\includegraphics[width=16cm,height=16cm,keepaspectratio]{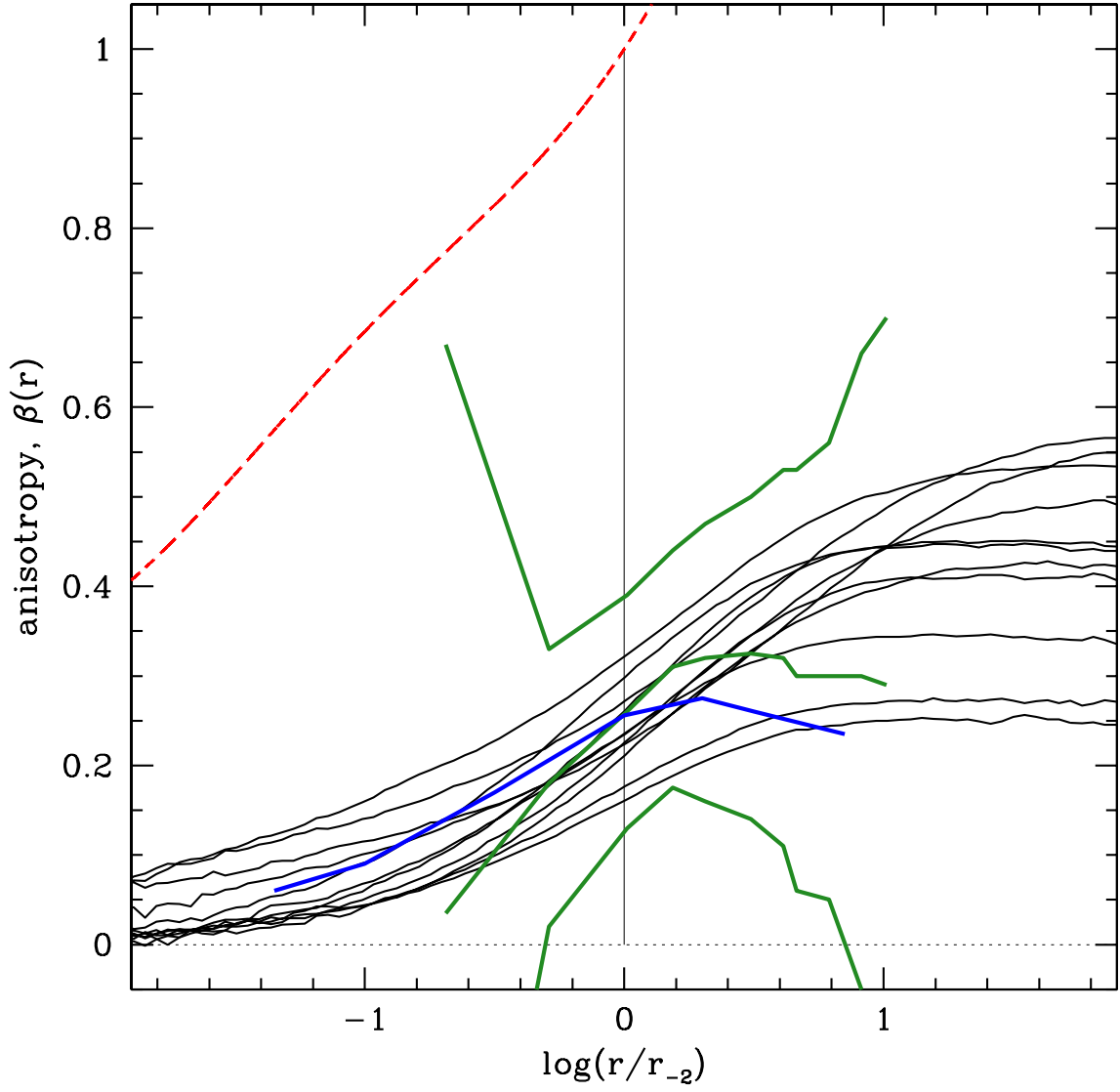}
\caption{Similar to Fig.~\ref{betaprof}. Here we compare our anisotropy profiles 
(black curves) with those from numerical simulations. The blue curve is the average profile 
from Fig.\ 3b of \citet{2011MNRAS.415.3895L}. The green curves are the average and upper and 
lower limits of relaxed systems taken from \citet{2012ApJ...752..141L} (blue curves in their Fig.\ 13).
Because their radius is in units of the virial radius while our systems do not have a defined 
virial radius,  we scaled their horizontal axis to have the same velocity anisotropy value at 
$r_{-2}$ as the \citet{2011MNRAS.415.3895L} data. A subset of our models that fit comfortably 
within the green curve bounds are shown in black.
}
\label{betaprofsim}
\end{figure}

We used density profiles for DARKexp $\phi_0=4$; other values of $\phi_0$ and 
other types of profiles give similar results. Each $L^2$ distribution results 
in a different anisotropy profile. Figure~\ref{betaprof} shows the anisotropy 
profiles of all the halos used in this paper. They span a wide range of 
possibilities. The subset of these profiles that are similar to those in cosmological
N-body $\Lambda$CDM simulations are shown as black curves in Figure~\ref{betaprofsim}.
The simulations are represented here by two recent works. The blue curve is the average 
profile from Fig.\ 3b of \citet{2011MNRAS.415.3895L}. The green curves are the average 
and upper and lower limits of relaxed systems taken from \citet{2012ApJ...752..141L} 
(blue curves in their Fig.\ 13). Because their radius is in units of the virial radius 
while our systems do not have a defined virial radius,  we scaled their horizontal axis 
to have the same velocity anisotropy value at $r_{-2}$ as the \citet{2011MNRAS.415.3895L} data.
This may not be the optimal scaling, but whatever scaling one adopts, it is clear that
different simulations do not completely agree with each other. This is also clear from the
examination of the seven $\beta(r)$ profiles presented in Fig.\ 11b of \citet{2010MNRAS.402...21N}.

In Figure~\ref{smintro} we show an example of one of our halos. The upper left 
panel shows one possible prescription for the distribution in $(L/L_{\rm{circ}})^2$ 
of particles binned by energy into six energy $\epsilon$ bins. The most bound 
particles (black and red histograms) show nearly uniform distribution in 
$L^2$. A perfectly uniform distribution would result in an isotropic system
\citep{2010ApJ...725..282W}.
The least bound particles (magenta and blue histograms) are biased toward 
radial orbits, and the near circular orbits are completely absent at these 
energies. The upper right panel has the full $N(E,L^2)$ distribution, shown 
with linear and log vertical axis. 
The magenta line represents circular orbits. The linear plot shows that
for the least bound particles, at $\epsilon\,\simgt\, 2$, near circular 
orbits are absent. The lower left panel has radial (main plot) and tangential 
(inset) VDF. We denote the radial VDF distribution by $N_{ur}(u_r)$, where 
$u_r=v_r/\sigma_r$, and $\sigma_r$ is the radial velocity dispersion at that 
radius. The area under each VDF is normalized to 1, 
$\int_{0}^{\infty}N_{ur}(u_r)\,du_r=1$. For all the halos we considered 
tangential VDFs peak at $u_t=0$. This is not so for radial VDF, which 
sometimes show deficits of orbits at small radial speeds. In this case the 
third radial bin from the center (blue histogram) shows a central `crater'.
The lower right panel shows the density profile, $\rho(r)$ as 
$\log(\rho r^2)$ (thick solid line), of DARKexp with $\phi_0=4$, and the 
anisotropy profile, $\beta(r)$ (dashed line). The thin solid line is the NFW 
profile shown here for comparison. The horizontal axis is in units of $r_{-2}$,
the radius where the logarithmic density slope, 
$d\ln \rho(r)/d\ln r$, is equal to $-2$.

\subsection{Characterizing the $L^2$ distributions}\label{charac}

To see how the distribution in $L^2$ is related to the VDF shape and 
$\beta(r)$, we need to characterize all three quantities with simple 
parameters. 

\begin{figure}
\includegraphics[width=16cm,height=16cm,keepaspectratio]{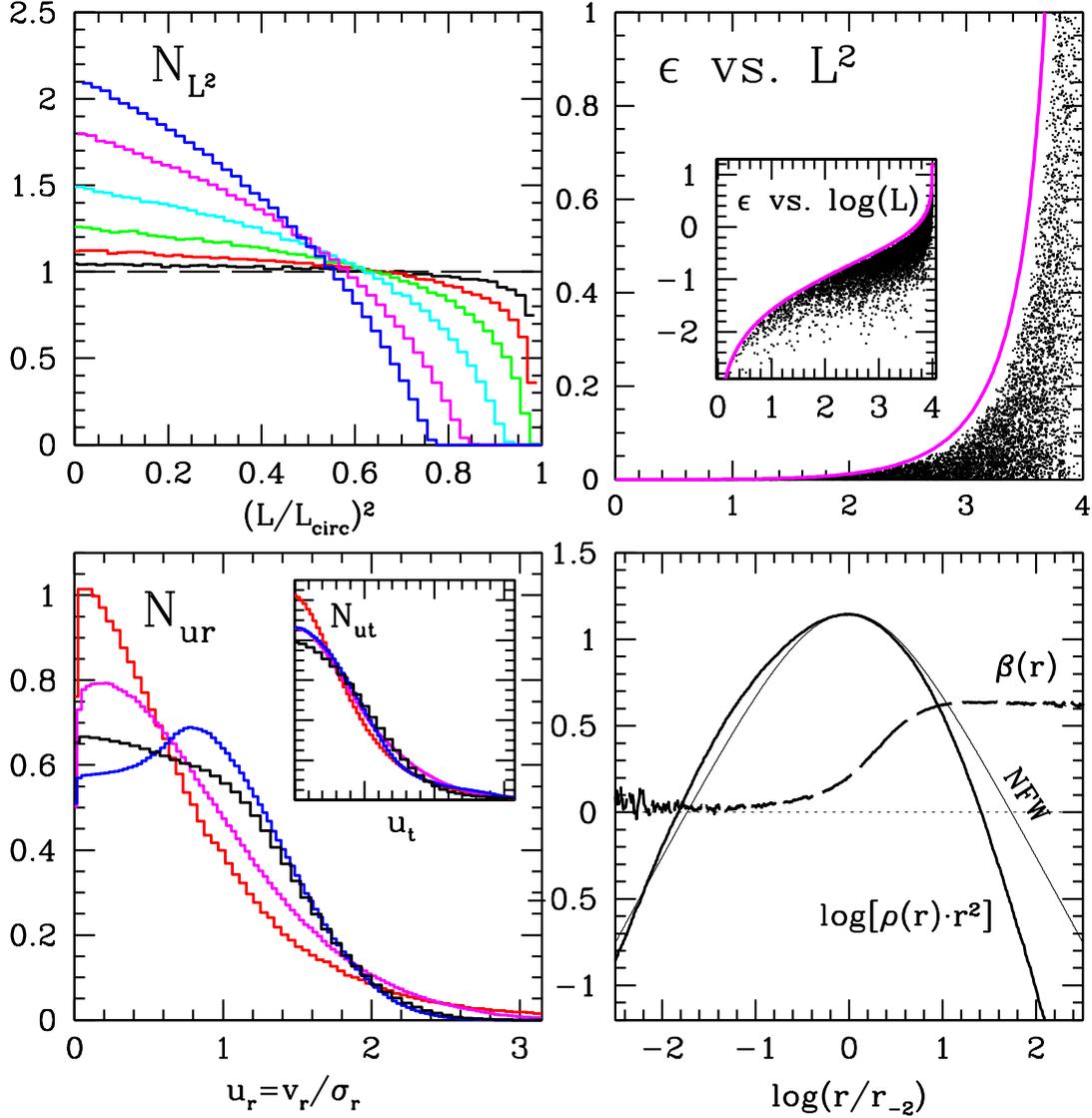}
\caption{An example of a halo. The halo is DARKexp with $\phi_0=4$.  
(Vertical axis labels are shown inside the individual panels.)
{\it Upper Left:}  Each curve is the distribution of particles in 
$\ell^2=(L/L_{\rm{circ}})^2$, where $L_{\rm{circ}}$ is the maximum angular momentum at 
that energy. The orbits are divided into six bins, of equal $\Delta \epsilon$.
In ascending order in energy (most to least bound particles), the six energy bins are 
black, red, green, cyan, magenta, and blue. 
{\it Upper Right:} The $N(E,L^2)$ distribution. The upper magenta envelope 
corresponds to circular orbits, $L_{\rm{circ}}$. The main plot shows $L^2$ plotted 
linearly. The inset shows $\log(L)$. Only a few hundred points are plotted,
whereas our halos have about $10^7$ particles each. 
{\it Lower Left:} The radial (main plot) and tangential (inset) velocity 
distribution functions for four radial intervals in the halo: red (inner 
most), black (outer most). 
{\it Lower Right:} Log of density times $r^2$. (Normalization is arbitrary.) 
NFW profile is shown for comparison, as a thin solid line. Anisotropy profile 
for the DARKexp halo is shown as a dashed line. The horizontal axis is radius 
in units of $r_{-2}$, where density profile has an instantaneous log-log 
slope of $-2$.
}
\label{smintro}
\end{figure}

We start with the $L^2$ distribution. For any given energy $\epsilon$, 
the distribution of particles in $\ell^2\equiv(L/L_{\rm{circ}})^2$ is denoted by 
$N_{L^2}(\ell^2)$, which is normalized, $\int_{0}^{1}N_{L^2}(\ell^2)\,d\ell^2=1$, 
where, by definition, $\ell^2$ runs from 0 to 1. We define the {\em curvature} of the 
$N_{L^2}(\ell^2)$ distribution for a given $\epsilon$ as 
\begin{equation}\label{Ccurv}
\!\!\tilde K(\epsilon)\!=\frac{\int\limits_{0}^{1}[N_{L^2}(\ell^2)-N_{L^2}(\ell_m^2)]\,d\ell^2}{[N_{L^2}(0)-N_{L^2}(\ell_m^2)]\,\ell_m^2/2}-\!1,\,\,K\!\!=\!\Bigl\langle\tilde K(\epsilon)\Bigr\rangle_{\!{{\rm all }\,\epsilon}}
\end{equation}
where $\ell_m$ is the largest $\ell$ where $N_{L^2}(\ell^2)$ is non-zero. In the upper 
left panel of Figure~\ref{smintro}, the distributions corresponding to the most bound 
particles (black and red histograms) have $\ell_m=1$, while the distribution corresponding
to the least bound particles (blue histogram) has 
$\ell_m^2=0.76$, and near circular orbits are completely absent. In words, 
Equation~(\ref{Ccurv}) is an expression for the curvature of the line 
connecting  the highest point of $N_{L^2}(\ell^2)$, i.e., at $\ell=0$ and the lowest 
point, $N_{L^2}(\ell^2=j_m^2)$. $K(\epsilon)$ is 0 for straight-line 
distributions, negative for convex (sagging) $N_{L^2}(\ell^2)$ distributions, and 
positive for concave (upward `bulging') $N_{L^2}(\ell^2)$ distributions. The 
curvatures of all six histograms in this figure are positive, $K>0$.

We define two more quantities for $N_{L^2}(\ell^2)$. The {\em average vertical extent} 
of the $N_{L^2}(\ell^2)$ histograms, 
\begin{equation}
\bar\Delta=\Bigl\langle N_{L^2}(\ell\!=\!0)-N_{L^2}(\ell\!=\!1)\Bigr\rangle_{{\rm all }\,\epsilon},
\end{equation}
and the {\em vertical rms dispersion} between the $N_{L^2}(\ell^2)$ histograms, which 
we call $\Delta_{\rm{rms}}$. It quantifies how spread out the histograms at various 
energies are. If the distributions in $N_{L^2}(\ell^2)$ coincide at all energies, 
then $\Delta_{\rm{rms}}=0$. For the halo shown in Figure~\ref{smintro}, 
$\bar\Delta=1.26$ and $\Delta_{\rm{rms}}=0.26$.

Next, we quantify the shape of the VDF. Because our ultimate goal is to 
separate plausible halos from implausible ones we are especially interested in 
the crater, which we consider unphysical. 
\citep[VDFs of N-body simulations are either flat-topped or peaked at small 
speeds, see][]{2010JCAP...02..030K,2012ApJ...756..100H}.
Let $u_{rm}$ be the radial speed where $N_{ur}(u_r)$ is the largest. For VDFs 
that peak at $u_r=0$, $u_{rm}=0$, but for those with a crater, $u_{rm}>0$. The 
{\em fractional area of the radial VDF crater} is
\begin{equation}\label{vdfshape}
\Upsilon=\int\limits_0^{u_{rm}}\,[N_{ur}(u_r\!=\!u_{rm})-N_{ur}(u_r)]\,du_r.
\end{equation}
In the case of Figure~\ref{smintro} the blue VDF in the lower left panel 
has $\Upsilon=0.058$.

Finally, we quantify the shape of the anisotropy profile, $\beta(r)$. We 
consider $\beta(r)$ to be realistic if it is either increasing or staying 
constant with radius, and nearly zero (i.e., isotropic) at very small radii. 
(We remove and do not consider systems with anisotropy profiles that are 
monotonically falling at all radii.) With that in mind, we define $\beta_0$ 
to be the value at $\log(r/r_{-2})=-1.5$ and $\beta_{\rm{min}}$ to be the minimum 
value attained in the radial range between $\log(r/r_{-2})=-1.5$ and $0$. 
$\beta_{\rm{min}}$ is not always the same as $\beta_0$ because some anisotropy 
profiles have minima between $\log(r/r_{-2})=-1.5$ and $0$, and then increase 
at larger $r$. We quantify the $\beta$ profile with $B=\beta_0+\beta_{\rm{min}}$. 
This definition is somewhat arbitrary, and other variants of $B$ can be 
adopted. The anisotropy profile shown in the lower right panel of 
Figure~\ref{smintro} has $B=0.02$.

\subsection{Relating the $L^2$ distributions to the halo velocity 
properties}\label{relating}

\begin{figure}
\includegraphics[width=16cm,height=16cm,keepaspectratio]{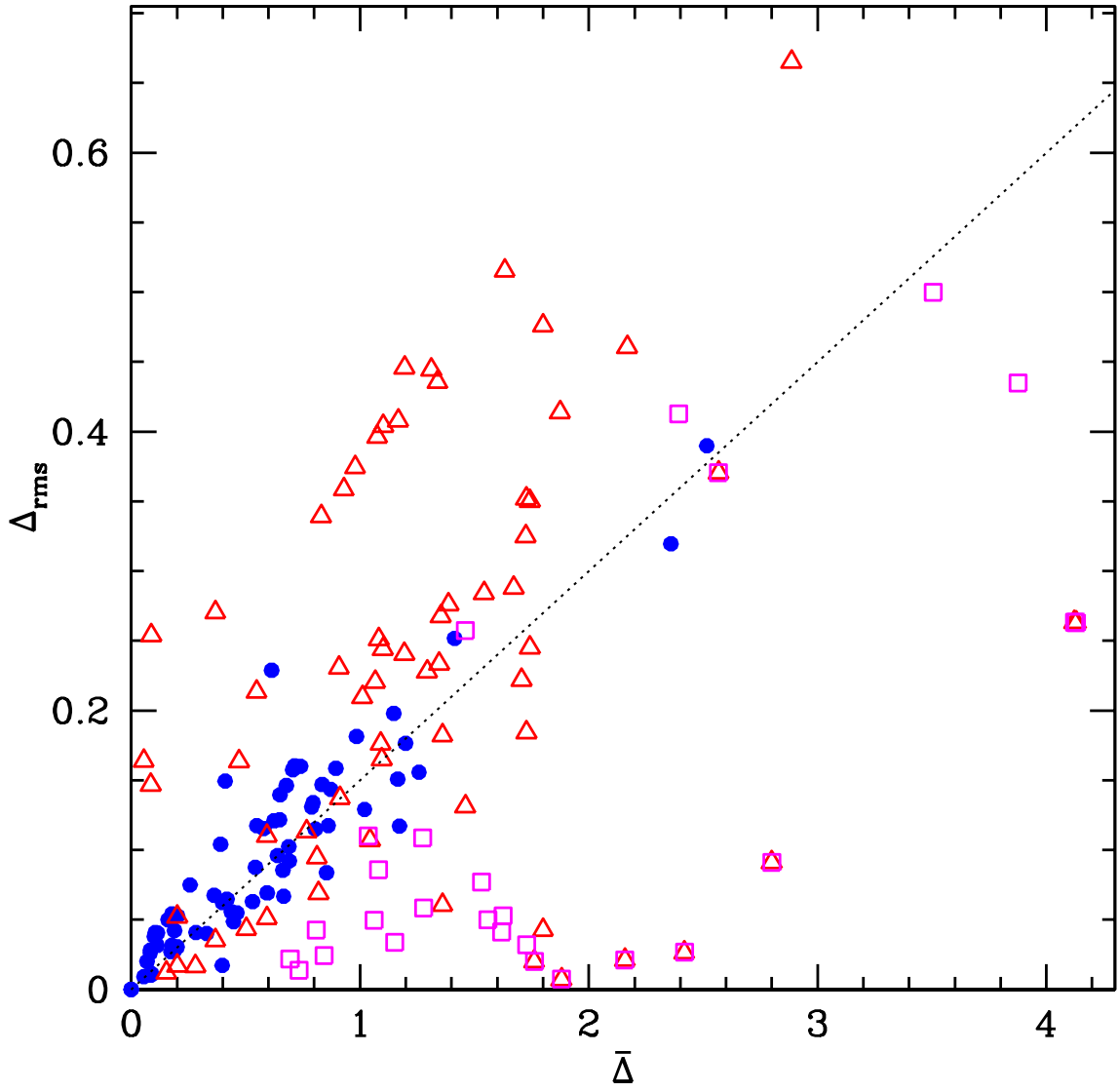}
\caption{$\bar\Delta$ vs. $\Delta_{\rm{rms}}$, both of which are determined from $N_{L^2}$. Red triangle points 
represent halos with VDF craters, $\Upsilon>0$. Magenta squares represent halos with $B>0.1$. Blue points 
gave $B<0.1$ and $\Upsilon<0$. The diagonal line
roughly separates the two. See Sections~\ref{charac} and \ref{relating} for details. }
\label{smsix}
\end{figure}

Having characterized $N_{L^2}$, the VDF, and $\beta(r)$ with simple 
parameters, we can now ask how the distribution of angular momentum, 
$N_{L^2}$, is reflected in the halos' VDF and $\beta(r)$.  In 
Figure~\ref{smsix} we plot $\bar\Delta$ vs.\ $\Delta_{\rm{rms}}$, both of which are 
determined from $N_{L^2}$. Red triangle points represent halos with VDF 
craters, $\Upsilon>0$. Magenta squares represent halos with $B>0.1$. Both of 
these types of systems are unrealistic. It is immediately obvious from the 
plot that these two types of systems tend to inhabit different parts of the 
plot, roughly separated by a straight diagonal line. Most of the systems 
close to the line are astrophysically realistic (blue filled dots). However, 
there are some systems with VDF craters that are mixed in (red triangles 
among blue dots). 

\begin{figure}
\includegraphics[width=16cm,height=16cm,keepaspectratio]{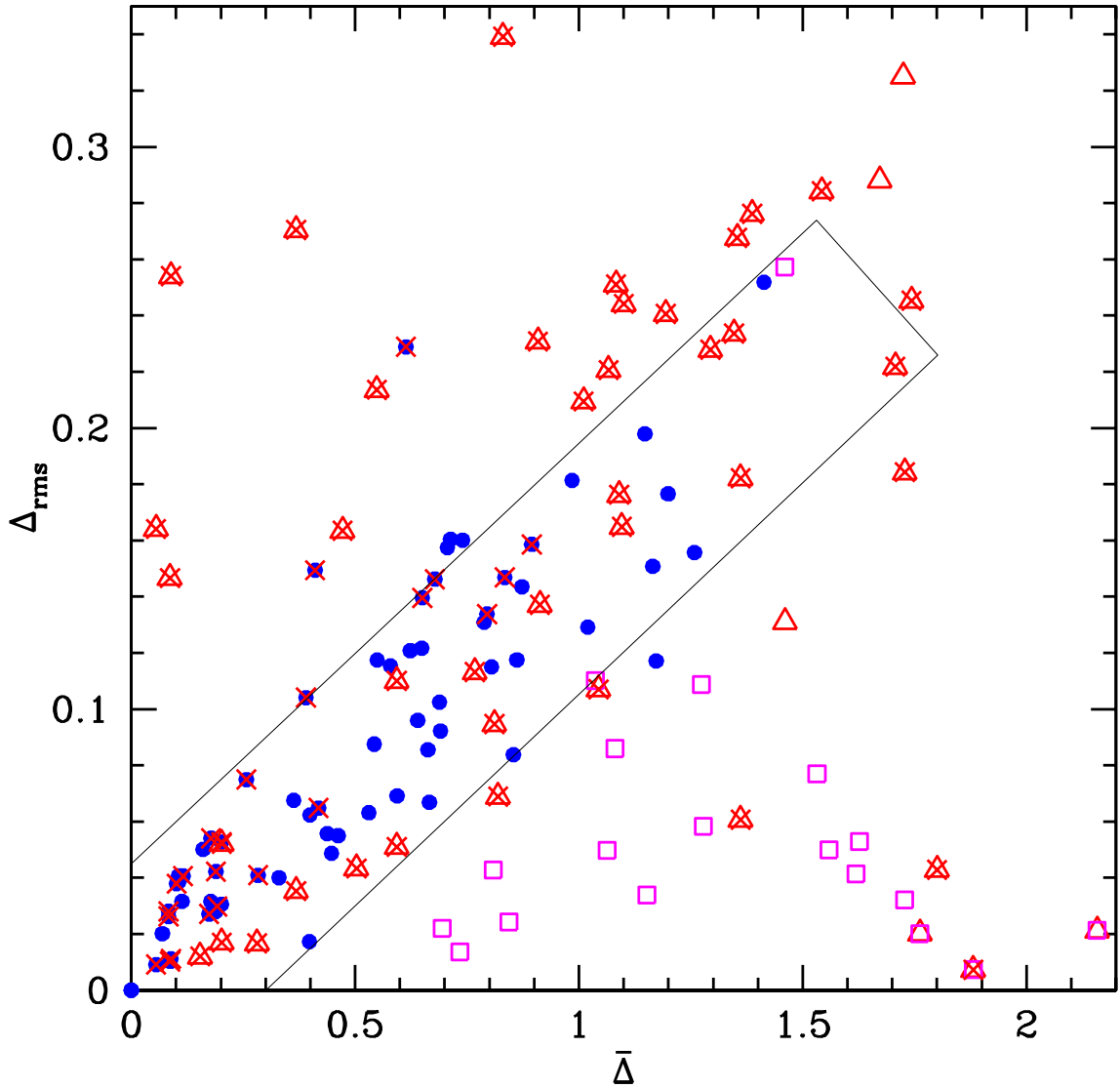}
\caption{Zoom-in on Figure~\ref{smsix}, now showing the systems with concave $N_{L^2}$ distributions ($K>0$) marked with
red crosses. The upper limits on $\bar\Delta$ and $\Delta_{\rm{rms}}$ do not extend as far as in Figure~\ref{smsix}; this
eliminates systems with $\beta(r)$ near 1 at large radii. The diagonal rectangular box delineates our selection criteria.
Systems that are inside the box and are not marked with a red cross have realistic VDF and $\beta(r)$ profiles.
See Sections~\ref{charac} and \ref{relating} for details.}
\label{smsixtwo}
\end{figure}

Figure~\ref{smsixtwo} shows only a portion of the Figure~\ref{smsix} diagram; 
it cuts out systems with large $\bar\Delta$ and $\Delta_{\rm{rms}}$, which have large 
anisotropies, close to $1$, at large radii.
In Figure~\ref{smsixtwo} we show that VDF crater systems ($\Upsilon>0$) can be 
identified and hence eliminated by using another property of the $N_{L^2}(\ell^2)$ 
distributions, namely the curvature $K$, defined by Equation~(\ref{Ccurv}). 
Points marked with red crosses have $K>0$. There is a very close 
correspondence between red triangles and red crosses. In other words, 
$N_{L^2}(\ell^2)$ distributions that are concave ($K>0$) almost always have VDF 
craters ($\Upsilon>0$), and systems that have VDF craters are almost always 
concave. 

Figure~\ref{smsixtwo} shows that 
convex $K<0$ systems restricted to lie in the box around the diagonal line have 
realistic VDF and $\beta(r)$ profiles. This selection uses $\bar\Delta$, 
$\Delta_{\rm{rms}}$ and $K$, i.e., it is based solely on the shape of 
$N_{L^2}(\ell^2)$. In Figure~\ref{smsixtwo} we have therefore accomplished our 
goal of isolating $N_{L^2}$ shapes that gives rise to realistic systems. 

\begin{figure}
\epsscale{1}
\includegraphics[width=16cm,height=16cm,keepaspectratio]{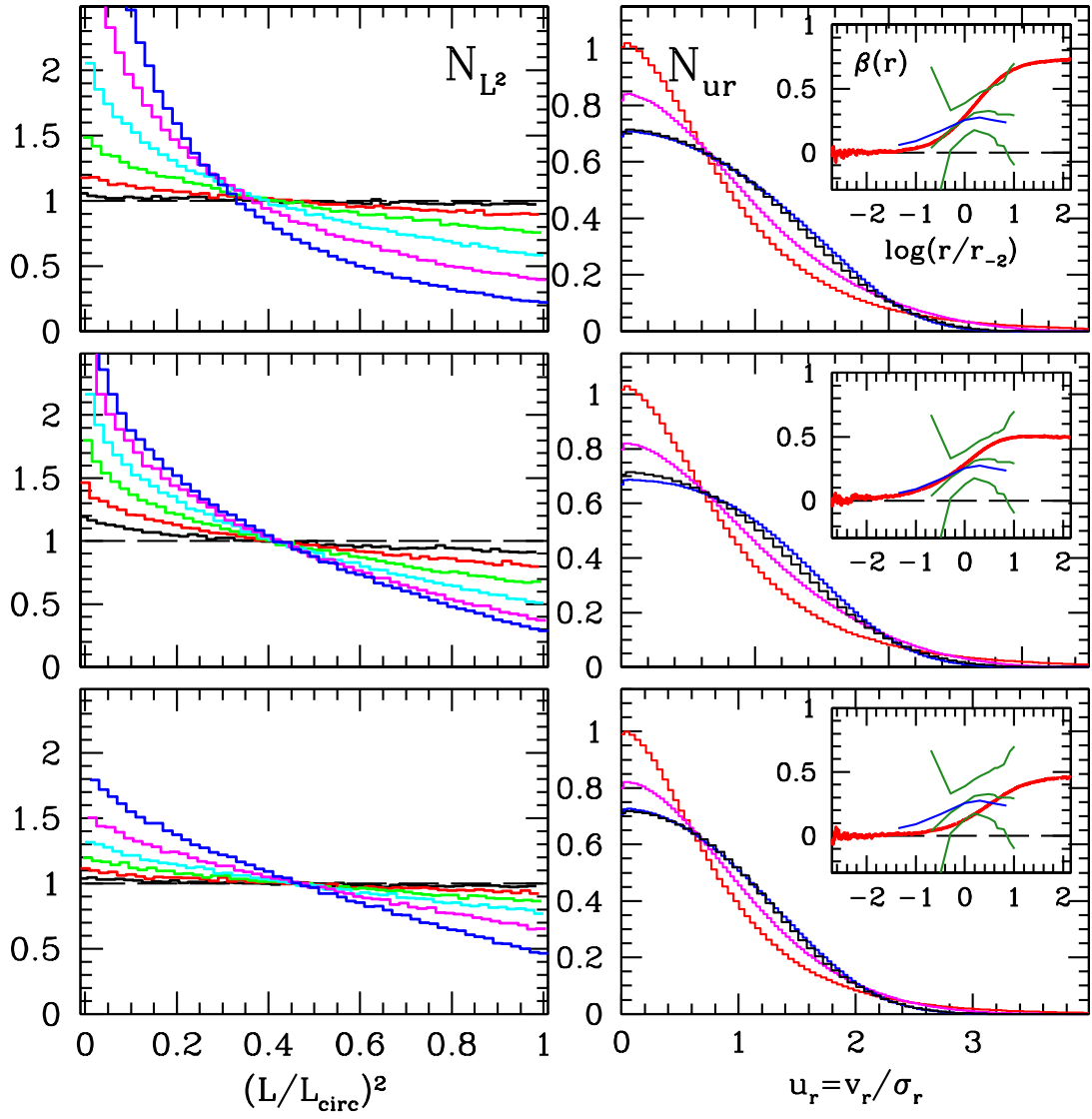}
\caption{Examples of the angular momentum distribution, VDF, and anisotropy profiles of three realistic systems.
The blue curve in the inset of the right panels is the average simulated velocity anisotropy profile 
from Fig.\ 3b of \citet{2011MNRAS.415.3895L}. The green curves are the average and upper and 
lower limits of relaxed systems taken from \citet{2012ApJ...752..141L}. See the caption of Figure~\ref{betaprofsim}
for details.
}
\label{smVDF4}
\end{figure}

Figure~\ref{smVDF4} shows three examples of realistic halos (blue dots inside 
the diagonal rectangle in Figure~\ref{smsixtwo}). Note that all three are 
isotropic at small radii, which is a consequence of $N_{L^2}$ being nearly 
flat at very large negative energies (black and red histograms in the left panels). For 
less negative energies the $N_{L^2}$ distributions favor low angular momentum orbits, 
but in such a way that $N_{L^2}$ is convex. The larger the $\bar\Delta$ the 
larger the velocity anisotropy at large $r$. These anisotropy profiles are 
not dissimilar to those seen in cosmological simulations, represented here by two
recent works,  \citet{2011MNRAS.415.3895L,2012ApJ...752..141L}; see Section~\ref{caseIII} for details.
   
\begin{figure}
\epsscale{1}
\includegraphics[width=16cm,height=16cm,keepaspectratio]{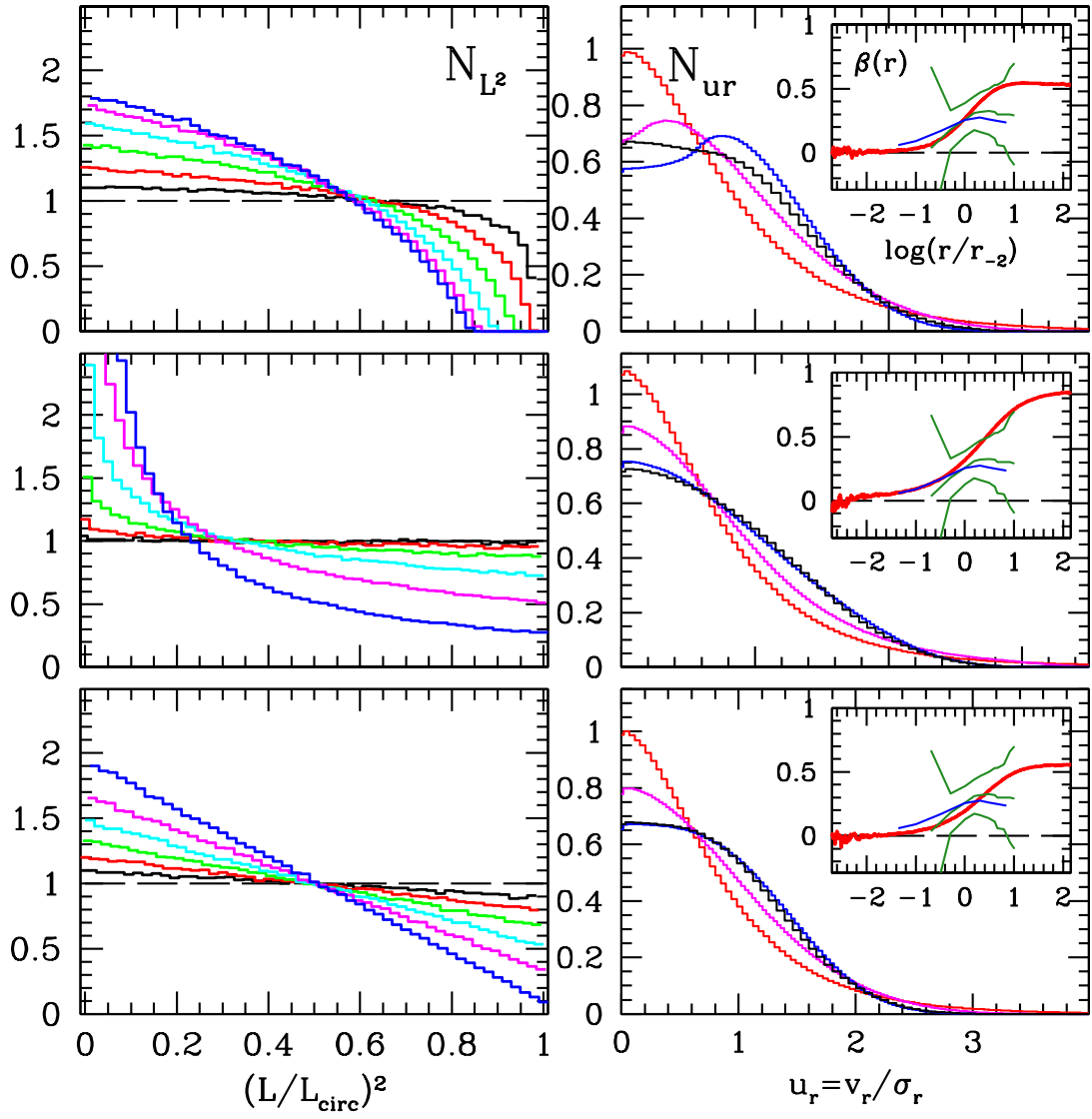}
\caption{Similar to Figure~\ref{smVDF4} for 3 systems inside the rectangular box of Figure~\ref{smsixtwo}. 
The top row shows a system that was correctly eliminated as unrealistic by our criteria, and for the other 
two systems our criteria fail; see Section~\ref{relating} for details.}
\label{smVDF5}
\end{figure}

Unfortunately, Figure~\ref{smsixtwo} does not provide a completely clean 
separation because the shapes of VDF and anisotropy profiles depend on the 
detailed properties of the $L^2$ distribution, which cannot be fully
captured with simple global parameters. 
In Figure~\ref{smVDF5} we show three more systems inside the diagonal 
rectangle of Figure~\ref{smsixtwo}. The top set of panels shows a system that 
is correctly eliminated by our criteria as unrealistic: its $N_{L^2}$ is 
concave, and it has VDF craters. The middle set of panels is a system for 
which our selection criteria fail by a small amount: the system's $B=0.106$, 
which is just outside our limit of $0.1$. It is represented by a magenta 
square in the upper portion of the diagonal rectangle. The bottom set of 
panels contains a system where our criteria fail, but in the opposite sense: 
they eliminate a realistic system because it has a slightly concave $N_{L^2}$ 
shape, $K>0$ (one of the blue dots with a red cross through it).

The bottom panels of Figures~\ref{smVDF4} and \ref{smVDF5} were generated 
using the Equation~(\ref{ex1}) prescription for distributing orbits in $L^2$, 
while the top panels of Figure~\ref{smVDF4} and the middle panels of 
Figure~\ref{smVDF5} were generated using Equation~(\ref{ex2}). 
Equation~(\ref{ex2}) (with constant $c\,\simgt\, 0.2$) is probably the best among 
the ones we tried in terms of generating systems that tend to lie mostly in 
the diagonal rectangle of realistic halos. Note that its dependence on 
$L^2$ is of exponential form $e^{-t(L^2)}$, similar to the exponential 
dependence on energy in the DARKexp $N(E)$. 

\begin{figure}
\includegraphics[width=16cm,height=16cm,keepaspectratio]{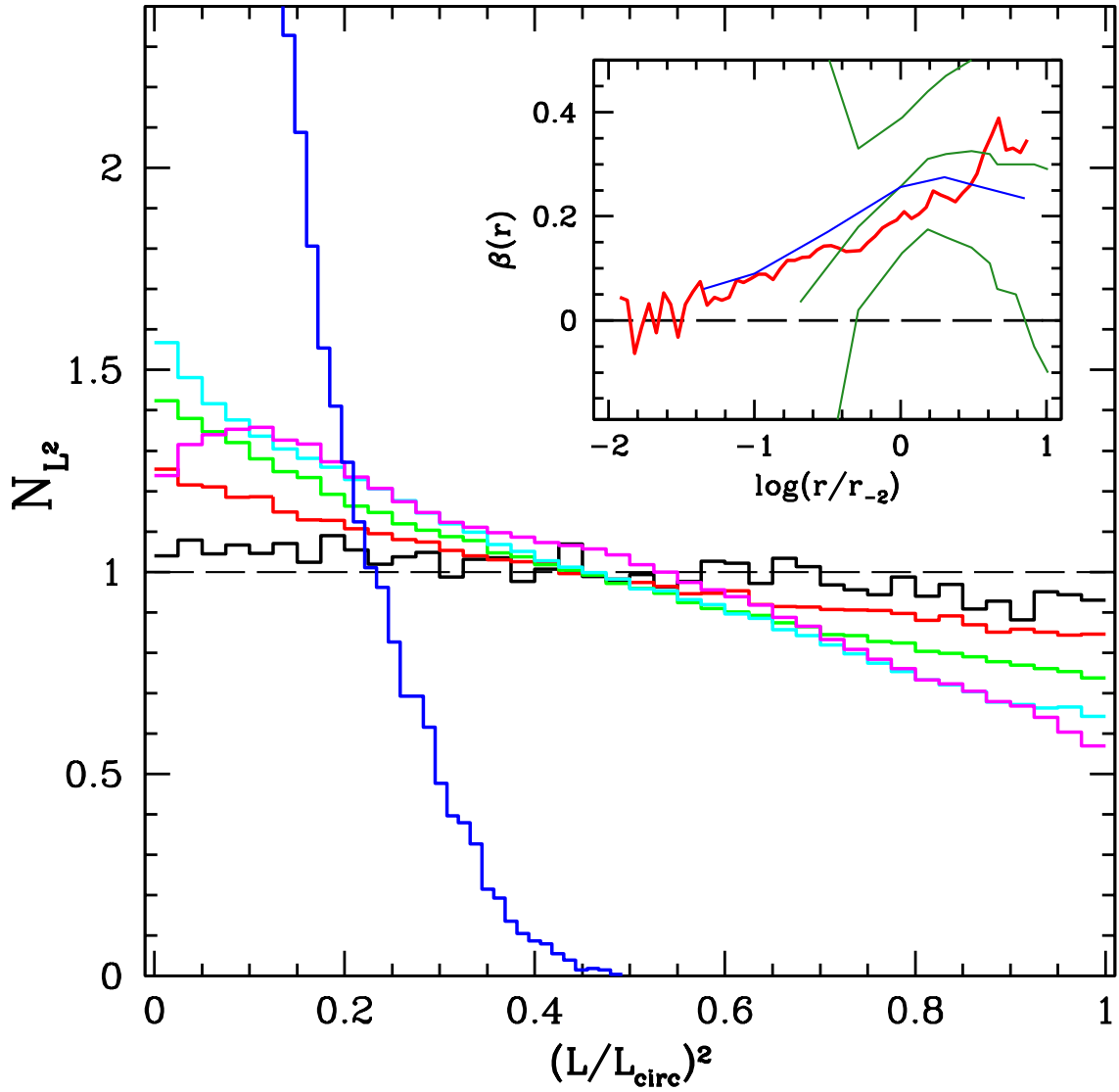}
\caption{The $(L/L_{\rm{circ}})^2$ distribution from N-body simulations of cluster-sized halos,
using the color scheme similar but identical to the one in other plots in this paper. 
The inset shows spherically averaged anisotropy profile. This is the
same set of halos that were shown to follow DARKexp $N(E)$ in 
\citetalias{2010ApJ...725..282W}. The blue curve in the inset of the right panel is the average 
simulated velocity anisotropy profile from Fig.\ 3b of \citet{2011MNRAS.415.3895L}. 
The green curves are the average and upper and  lower limits of relaxed systems taken from 
\citet{2012ApJ...752..141L}. See the caption of Figure~\ref{betaprofsim} for details.
}
\label{radek}
\end{figure}

The systems in Figure~\ref{smVDF4} can be compared qualitatively with typical 
halos from N-body simulations, shown in Figure~\ref{radek}. This is based on a 
sample of 36 cluster-size relaxed halos extracted from a $z\!=\!0$ snapshot of 
an N-body simulation of a standard $\Lambda$CDM cosmological model 
\citep[for details of the simulation and the halo catalog, see][]{2008MNRAS.388..815W}. 
This halo sample has already been used for calculating the six-dimensional distribution 
function as a function of energy and angular momentum, and testing its phenomenological 
model with radially changing anisotropy \citep{2008MNRAS.388..815W}. Each halo 
contains from $5\times 10^5$ to $5\times 10^6$ particles inside its virial 
sphere defined in terms of the mean overdensity, 
$\langle\rho\rangle/\rho_{crit}\approx 100$, where $\rho_{crit}$ is the 
present critical density. The energy distribution of these halos was also 
used in \citetalias{2010ApJ...725..282W}.

We conclude that realistic systems are characterized by two main properties of 
their $N_{L^2}$:
(1) $N_{L^2}$ shapes are straight or convex ($K\leq 0$), and 
(2) the dispersion in $N_{L^2}$ for different energies ($\Delta_{\rm{rms}}$) is 
proportional to the average vertical extent of the $N_{L^2}$ distributions
($\bar\Delta$), i.e., the systems lie inside the diagonal rectangular box of 
Figure~\ref{smsixtwo}. 


\section{Summary and conclusions}\label{secIV}

While DARKexp $N(E)$ was arrived at through statistical mechanics maximum 
entropy analysis, the extension to $N(E,L^2)$ considered in this paper is not. 
Instead, our starting premise was Equation~(\ref{NII}), which is the integral 
equation for the isotropic DARKexp, and is based on the 
assumption that obtaining the distribution of particles in $E$ is different 
from that in $L^2$. The two different treatments---maximum entropy for $E$ 
and a phenomenological approach for $L^2$---are appropriate because, as we 
argue in Section~\ref{genL2}, self-gravitating collapsing systems are free 
to redistribute their particles in $E$, but not in $L^2$. Because the 
redistribution of angular momentum is not unrestricted, not all portions of 
the $L^2$ space are equally accessible, making maximum entropy arguments
inappropriate.



By investingating the properties of a large number of halos all consistent
with Equation~(\ref{NII}) we conclude that astrophysically realistic halos 
must have $N([L/L_{\rm{circ}}]^2)$ distributions (for each energy separately) that 
are linear or somewhat convex in $(L/L_{\rm{circ}})^2$. Also, the 
$N([L/L_{\rm{circ}}]^2)$ distribution for most bound particles (largest negative energies) 
must be uniform, and become more tilted in favor of radial orbits for less bound particles. 
We give two examples of a prescription for $N(E,L^2)$ that generates realistic systems:
Equations~(\ref{ex1}) and (\ref{ex2}). The approximate constant ranges for 
Equation~(\ref{ex1}) are: $1\!<\!a\!<\!2$, $0.5\!<\!b\!<\!1.5$, and $0.5\!<\!c\!<\!1.5$, and for
Equation~(\ref{ex2}): $1\!<\!a\!<\!3$, $0\!<\!b\!<\!1.5$, and $c\!>\!0.2$. 

What can give rise to such $N([L/L_{\rm{circ}}]^2)$ distributions? We argued above
that because the particles are not well mixed in $L^2$, the distribution in $L^2$ 
probably depends on the details of the initial conditions and dynamics of halo collapse,
like radial orbit, and other instabilities, and is possibly somewhat different for 
isolated vs. cosmological collapses. The advantage of our approach is that because
we considered all astrophysically realistic systems, it encompasses both
of these, as well as other possible cases. Our next step is to use the form of 
$N([L/L_{\rm{circ}}]^2)$ obtained here to generate the distribution function $f(E,L^2)$, 
which can be compared to $f(E,L^2)$ measured from N-body simulations 
\citep[e.g.,][]{2008MNRAS.388..815W}.



\appendix

\section{Appendix A}\label{appA}

Suppose some $L$-distribution is conserved as a part of a statistical 
mechanical approach, which also conserves total mass and energy. The 
distribution in energy and angular momentum that corresponds to the most 
likely state is given by
\begin{equation}\label{DL}
N(E,L^2)\propto \exp(\tilde\beta\Phi_0-\tilde\beta\,E-\gamma\,L^\xi)-1.
\end{equation}
This is analogous to Equation~(30) of \citet{1994ApJ...424..106H}. We show 
below that the density profiles that correspond to Equation~(\ref{DL}) are 
not the same as for DARKexp, and that the velocity anisotropy profiles become 
more tangential with increasing radius. Thus these systems are very different
from those seen in numerical simulations.

\begin{figure}
\includegraphics[width=16cm,height=16cm,keepaspectratio]{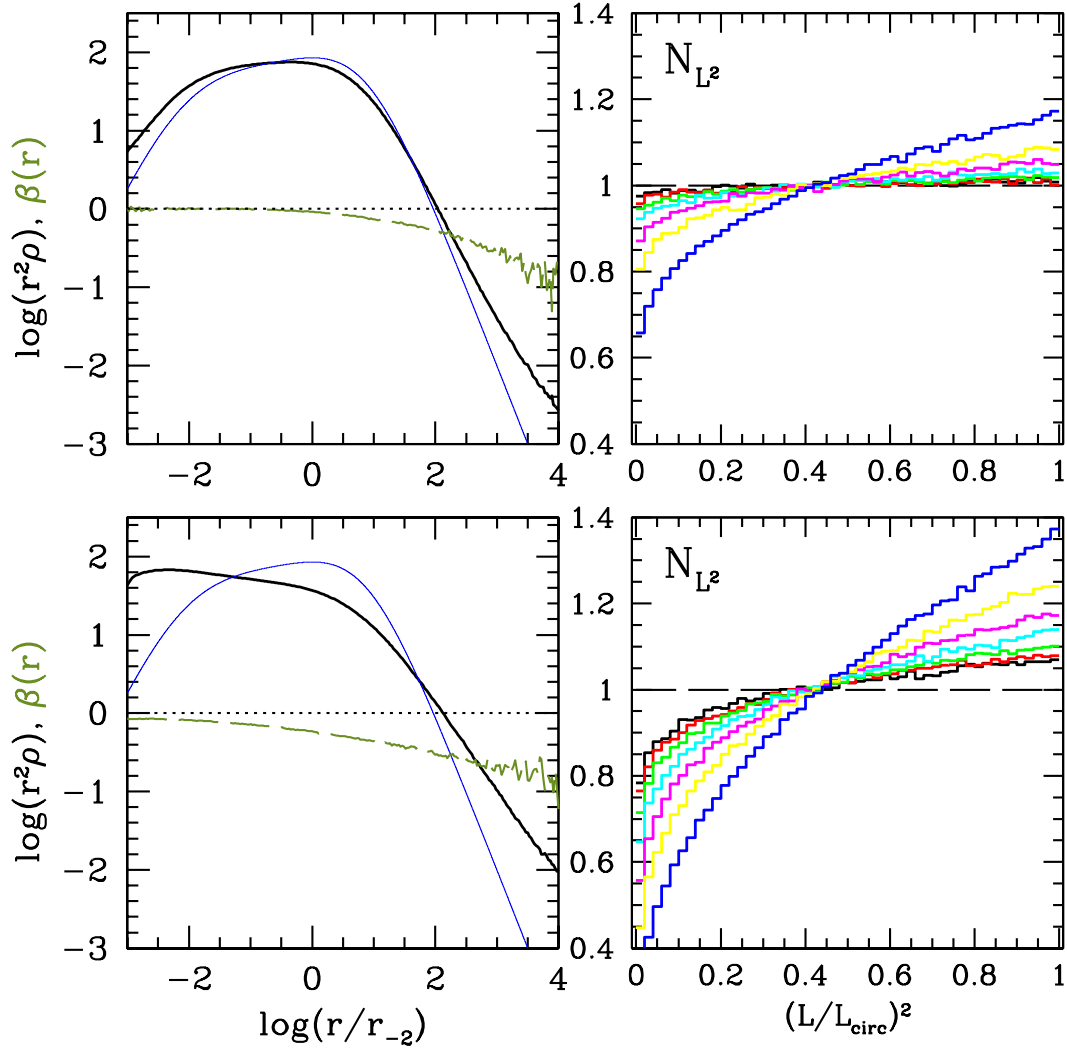}
\caption{Examples of two systems obeying Equation~\ref{NI}, with $\gamma=-1$, $\xi=0.5$,
$\phi_0=6$ (top panels) and $\gamma=-5$, $\xi=0.2$, $\phi_0=6$ (bottom panels). The
density profiles are multiplied by $r^2$ and plotted in the left panels as thick
black curves; the blue thin lines are DARKexp with $\phi_0=6$ shown for comparison. 
The horizontal axis is in units of $r_{-2}$ of the corresponding DARKexp profile. 
Right panels present the distribution of $L^2$, which is to be compared to those in 
Figure~\ref{smVDF4}. The distributions of $L^2$ in the present figure are biased 
towards high angular momenta, which is also reflected in the anisotropy profile 
(olive-green long-dash line in left panels).
}
\label{sm6L14a}
\end{figure}

We note that not all ($\gamma$, $\xi$) parameter combinations produce density
profiles. Systems with $\gamma>0$ or $\xi<0$ did not converge, and neither did 
systems with $\gamma\ll -10$ or $\xi\gg 1$. In Figure~\ref{sm6L14a} we show two systems 
that did converge. Their parameters are $\gamma=-1$, $\xi=0.5$, $\phi_0=6$ 
(top panels), and $\gamma=-5$, $\xi=0.2$, $\phi_0=6$. Other systems have similar 
general characteristics. 

\section{Appendix B}\label{appB}

Here we consider a special sub-class of systems described by 
Equation~(\ref{NI}), which produce analytical solutions.

The $N(E,L^2)$ distribution can be represented in the $E$ vs.\ $L^2$ plane. 
For every $E$ there is a maximum $L=L_{\rm{circ}}=L_c$ that corresponds to the 
circular orbit of that energy. The set of these $L_c$ forms an 'upper' 
envelope in the $E$ vs.\ $L^2$ plane. In a general case the density of 
particles along that envelope will vary as a function of $E$. Let us suppose 
that there is a sub-class of systems where the density is constant. Then 
Equation~(\ref{DL}) reduces to $N(E_c,L_c^2)=$const.\ on that envelope. 
Equivalently, $\tilde\beta\Phi_0-\tilde\beta\,E_c-\gamma\,L_c^\xi=$const. 
Differentiating, we get
\begin{equation}\label{envcond}
\frac{dE_c}{dL_c^\xi}=-\frac{\gamma}{\tilde\beta}={\rm const},
\end{equation}
because $\gamma$ and $\tilde\beta$ are constants for a given system. The left 
hand side of Equation~(\ref{envcond}) can be expressed differently, as it refers 
to circular orbits. Circular speed $v_c$ is given by $v_c^2=r d\Phi/dr$, and the 
corresponding energy is $E_c=\frac{1}{2}v_c^2+\Phi$. Using these we get 
$E_c=\frac{1}{2}r\Phi^\prime+\Phi$, and $L_c^2=r^3\Phi^\prime$, where primes 
denote differentiation with respect to radius. Next, we can obtain expressions 
for $dE_c/dr$ and $dL_c^\xi/dr$, and hence
\begin{equation}\label{circcond}
\frac{dE_c}{dL_c^\xi}=\frac{1}{\xi r^2}\Bigl(r^3\Phi^\prime\Bigr)^{1-(\xi/2)}.
\end{equation}
Combining Equations~(\ref{envcond}) and (\ref{circcond}) we get
\begin{equation}\label{phigradME}
\Phi^\prime(r)=\Bigl(-\frac{\tilde\beta}{\gamma\xi}\Bigr)^{2/(\xi-2)}\,r^{(2-3\xi)/(\xi-2)}.
\end{equation}
Integrating, we have
\begin{equation}\label{phiME}
\Phi(r)=\Bigl(-\frac{\tilde\beta}{\gamma\xi}\Bigr)^{2/(\xi-2)}\,\,\frac{(2-\xi)}{2\xi}\,\,r^{2\xi/(2-\xi)}+C,
\end{equation}
and applying the Poisson equation finally gives Equation~(\ref{rhoME}):
\begin{equation}\label{rhoME2}
\rho(r)=\frac{1}{4\pi}\Bigl(-\frac{\tilde\beta}{\gamma\xi}\Bigr)^{2/(\xi-2)}\Bigl(\frac{2+\xi}{2-\xi}\Bigr)\,r^{-4(\xi-1)/(\xi-2)}.
\end{equation}

Not all combinations of parameters $\tilde\beta$, $\gamma$, $\xi$ are allowed.
The factor $[-\tilde\beta/(\gamma\xi)]^{2/(\xi-2)}$ in the above
equations tells us that $\tilde\beta/(\gamma\xi)$ has to be negative. 
There are four ways of realizing this, and we discuss these cases separately.
In {\bf Ia} and {\bf Ib}, the inverse temperature is negative, $\tilde\beta<0$,
as in DARKexp, but in  {\bf IIa} and {\bf IIb}, it is positive.
Note that all four cases guarantee that $dE_c/dL_c>0$, i.e. that as the energy of a circular
orbit is increasing, its angular momentum must increase as well. On the other hand, 
Equation~(\ref{circcond}) can have either sign.

{\bf Case Ia:} $\tilde\beta<0$, $\gamma>0$ and $\xi>0$.\\
In Equation~(\ref{DL}) the term $\gamma\,L^\xi$ is independent of energy,
so for $\tilde\beta E\rightarrow\tilde\beta\Phi_0$, $N(E,L^2)$ can get as small as $-1$, i.e. it
can become negative. This is not allowed, and so this combination of parameters is
ruled out.

{\bf Case Ib:} $\tilde\beta<0$, $\gamma<0$ and $\xi<0$.\\
$N(E,L^2)$ cannot become negative, so in general, such solutions are allowed.
Looking at the factor $(2+\xi)/(2-\xi)$ in Equation~(\ref{rhoME2}) we see that 
only $-2<\xi<0$ are allowed. The factor $(2-\xi)/(2\xi)$ in Equation~(\ref{phiME}) guarantees
that the potential is negative, so constant $C$ can be set to zero.
The two limiting solutions are:
as $\xi\rightarrow 0$, $\rho(r)\rightarrow r^{-2}$, and 
as $\xi\rightarrow -2$, $\rho(r)\rightarrow r^{-3}$.

{\bf Case IIa:} $\tilde\beta>0$, $\gamma<0$ and $\xi>0$.\\
$N(E,L^2)$ in Equation~(\ref{DL}) can become negative, especially when $E\rightarrow 0$,
and $-\gamma L^\xi$ cannot compensate for negative $\tilde\beta \Phi_0$. Therefore this 
combination of parameters is ruled out.

{\bf Case IIb:} $\tilde\beta>0$, $\gamma>0$ and $\xi<0$.\\
$N(E,L^2)$ is always negative, so this combination of parameters is not allowed.


We conclude that if $L^\xi$ is treated the same way as total mass and total energy, 
then maximizing entropy subject to condition of Equation~(\ref{envcond}) gives potential 
and density profiles that are power laws in radius, with the values of constants 
$\tilde\beta$, $\gamma$, $\xi$ given by {\bf Case Ib}, and the allowed density profile slopes
spanning the range from $-2$ to $-3$. Just like in Appendix~A, we conclude that the
solutions obtained here are very different from DARKexp, and very different from
systems obtained in numerical simulations.



It is interesting to note that Equation~(\ref{DL}) does not reduce to DARKexp 
for isotropic systems. When $\xi\rightarrow 0$, Equation~(\ref{DL}) becomes 
$N(E,L^2)\propto \exp(\tilde\beta\Phi_0-\tilde\beta\,E-\gamma)-1.$  Because $\gamma$ cannot 
be $0$ (see Equation~\ref{envcond}), the $N(E,L^2)$ distribution does not reduce 
to DARKexp for any allowed parameter values.

\acknowledgments
The Dark Cosmology Centre (DARK) is funded by the Danish National Research 
Foundation. L.L.R.W. would like to thank DARK for their hospitality. The authors 
are grateful to Stefan Gottl\"ober, who kindly allowed one of his CLUES simulations,\\
({\tt {http://www.clues-project.org/simulations.html}}) to be used in this paper. 
The simulation has been performed at the Leibniz Rechenzentrum (LRZ) Munich.

\vskip0.15in
\bibliography{smo4}

\end{document}